\documentclass[aps,twocolumn,prl,tightenlines,floatfix,showpacs]{revtex4}
\usepackage[dvips]{graphicx}
\usepackage[english]{babel}
\usepackage{amsmath}
\usepackage{amssymb}
\usepackage{slashed}

\newcommand{\vk}{\mathbf{k}}

\newcommand{\vq}{\mathbf{q}}
\newcommand{\sumk}{\sum_{\mathbf{k}}}

\newcommand{\be}{\begin{eqnarray}}
\newcommand{\ee}{\end{eqnarray}}
\newcommand{\p}{\partial}

\begin{document}

\title{Theory of Fluctuating Charge Ordering in the Pseudogap Phase of the Cuprates Via A Preformed Pair Approach}

\author{Yan He, Peter Scherpelz, and K. Levin}

\affiliation{James Franck Institute and Department of Physics,
University of Chicago, Chicago, Illinois 60637, USA}

\date{\today}

\begin{abstract}
We study the static and dynamic behavior of
charge ordering within a $d$-wave pair pseudogap
(pg) scenario.
This is addressed using a density-density correlation
function derived from the
standard pg self energy, $\Sigma$ and
compatible with the longitudinal and
transverse sum rules. The
broadening
factor $\gamma$ in $\Sigma$ reflects the breaking of pairs into
constituent fermions.
We
apply this form for $\Sigma$ (derived elsewhere for high fields)
to demonstrate the existence of
quantum oscillations in a
non-Fermi liquid pg state.
Our conclusion is that the
pseudogap-induced pairbreaking, via $\gamma$,
allows the underlying fermiology to be revealed;
in YBCO, finite $\omega$ and $\gamma$ enable antinodal
fluctuations, despite the competition with a $d$-wave
gap in the static and superconducting limits.
\end{abstract}

\maketitle

\textit{Introduction}
One of the most exciting developments in the field of high temperature
superconductivity has arisen recently with the growing evidence for
charge ordered states
\cite{Sawatzky13,RSXS,Hudson13,Basov10,Keimer09}
now evident for
a range of hole concentrations in
the underdoped regime.
Some of the earliest indications for this
charge ordering were associated with reconstructed Fermi surfaces
inferred from quantum oscillations \cite{Taillefer07}.
While static charge ordering signatures appear most
clearly at these high magnetic fields \cite{Hardy13}, there is evidence that even
in zero field there is a fluctuating or dynamic propensity
\cite{Bonn13,Sawatzky13} for
the same charge ordering.
Central to these observations is the uncertainty over the
charge ordering wave-vector,
which may differ in different cuprate families. There
are claims that it is associated with both
nodal nesting (NN) as well as
anti-nodal (AN) nesting \cite{Boebinger11,Taillefer07,Sebastian11}.
(Here the nomenclature reflects the
nodal/antinodal anisotropy inherent in a $d$-wave order parameter.)
%
It could be argued that a discovery of this new form
of order in the cuprates presents evidence against a preformed pair
interpretation of the pseudogap.
Also problematic for a preformed
pair scenario is the growing evidence \cite{Hudson13,Sebastian11}
that the charge ordering
is \textit{antinodal}, since the
same $\mathbf{k}$ states participate in both the $d$-wave pseudogap
and the AN ordering.

In this paper, because of its importance, we look more deeply into
charge ordering within
a scenario in which the pseudogap derives
from
$d$-wave preformed pairs. We do so by calculating
the
associated density-density correlation functions, $P_{\rho,\rho}(\mathbf{q}, \omega)$
and
demonstrate how dynamic charge ordering fluctuations are
a reflection of the underlying fermiology in the presence of a gap.
Importantly, our work begins with the widely accepted
\cite{Malypapers,Normanphenom}
form of zero field self energy
which is known to give rise to Fermi arcs (with bandstructure
$\xi_{\mathbf{p}}$) 
\be
\Sigma(P)=- i \gamma' -\Delta_{pg}^2G_0^{\gamma}(-P)\equiv -i \gamma' + \frac{\Delta_{pg}^2}{i\epsilon+\xi_{\mathbf{p}}
+ i \gamma}.
\label{eq:1}
\ee

A key contribution of this paper is that we establish the
form of the density-density correlation function associated with
this self energy, in a manner
analytically consistent with the longitudinal and transverse sum rules.
Using this
we investigate the zero field, $H=0$, possible instabilities
(dynamic and static) in the presence of a $d$-wave pseudogap.
Depending on the fermiology we find both nodal and anti-nodal dynamic charge
ordering tendencies. Our work emphasizes the latter.
Coexistence of (albeit, dynamic) anti-nodal charge ordering and the pseudogap is shown
to derive from the
``pairbreaking" contribution (associated with $\gamma$ in Eq.~(\ref{eq:1}))
to the
density-density correlation
function.
This pair breaking dominates the quasi-particle scattering
(or nesting) contribution to the spectral weight at low $T$.
Stated alternatively, pairs need to be broken into their composite
fermions in order to contribute to the charge correlation
function.
Finite frequency enables this pairbreaking.
Since $\gamma$ is necessarily absent in the superconducting
self energy where the condensate pairs are infinitely
long lived, we conclude that
the pseudogap (with $\gamma \neq 0$) 
plays an important role in enabling
antinodal charge fluctuations.

We secondarily address the implications of this
self energy (Eq.~(\ref{eq:1})) for quantum oscillation experiments.
We show
that oscillatory behavior is found in thermodynamics for this non-Fermi liquid
pseudogap phase,
due primarily to the pairbreaking
associated with $\gamma$.
In this paper we include this study because of its
relevance to charge ordering and to counter the belief that
such oscillations imply Fermi liquid behavior. It should be stressed, however,
that this paper is otherwise devoted to $H=0$ behavior.
In earlier work we have shown \cite{Peter2} using
Gor'kov theory that the same self energy applies to the very high field limit.

Our approach can
be compared with others in the literature \cite{SenthilLee,Randeria13}
where it is claimed that quantum oscillations are a signature of
a high field Fermi liquid state,
and argue for a three peaked spectral function \cite{Micklitz}.

\textit{Theory} We next show how the self energy
in Eq.~(\ref{eq:1}) leads to a form for the diamagnetic current
$\frac{\tensor{n}}{m}$
which can be used to
make an ansatz for the
current-current correlation function; from this one can
readily infer the density-density
correlation function in the pseudogap phase.
From the definition of $\frac{\tensor{n}}{m}$
and by integration by parts, we can write the diamagnetic
contribution in terms of the full Green's function $G(P)$ as
\begin{eqnarray}
\label{eq:2}
\frac{\tensor{n}}{m}&=&2\sum_K\frac{\tensor{1}}{m}G(K)
=2\sum_K\frac{\partial^2\xi_{\mathbf{k}}}{\partial\mathbf{k}\partial\mathbf{k}}G(K)\nonumber \\
&=&2\sum_KG^2(K)\frac{\partial\xi_{\mathbf{k}}}
{\partial\mathbf{k}}\frac{\partial\xi_{\mathbf{k}}}{\partial\mathbf{k}}\big(1
-\Delta_{pg}^2(G_0^{\gamma})^2(-K)\big).
\end{eqnarray}
Here $K=(i\omega_n,\vk)$ and $\sum_K=T\sum_n\sum_{\vk}$.

\begin{figure}
\includegraphics[width=2.9in,clip]{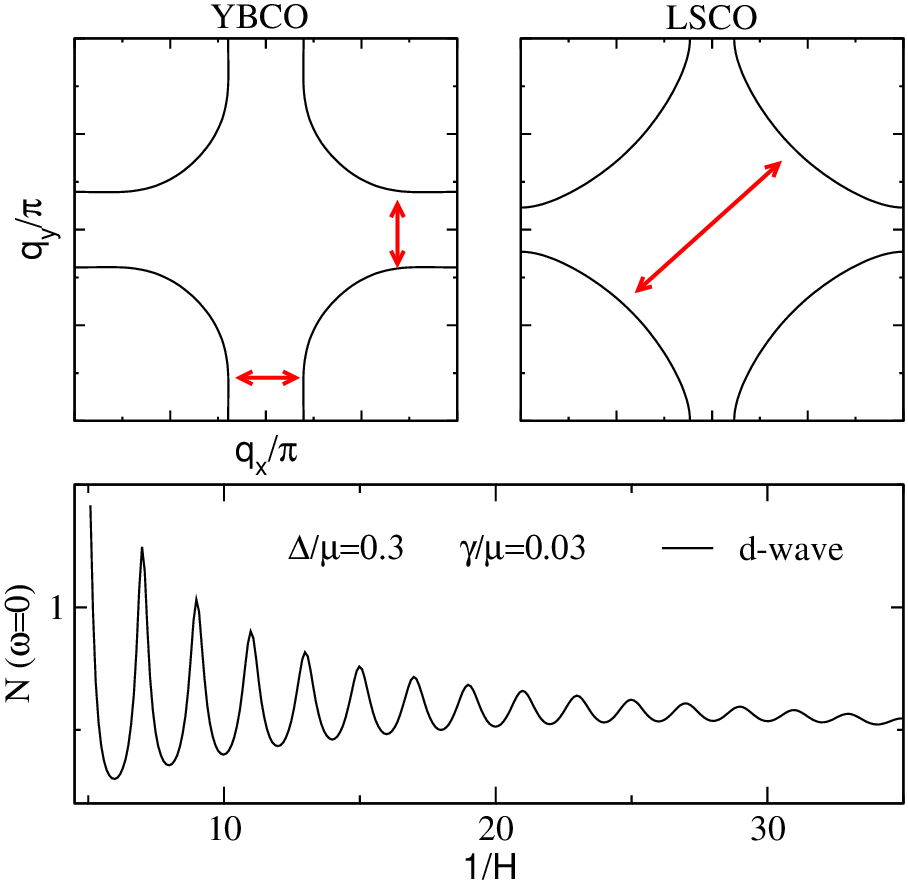}
\caption{Fermi surfaces for YBCO and LSCO with arrows indicating
the dominant nesting vectors.
The lower panel shows quantum oscillations persist in
a non-Fermi liquid pseudogap state, due to the
finite pair lifetime reflected in $\gamma^{-1}$. Their amplitude is reduced by
a factor of about 5 from the standard Lifshitz-Kosevich theory.}
\label{FS}
\end{figure}

Now we use the constraint that there is no Meissner effect in the
normal state to first determine the current-current
correlation function at four-vector $Q=0$, $P_{JJ}(0)$, and
then reconstruct $P_{JJ}(Q)$.
This latter ansatz however will be tested against two sum rules.
Using Eq.~(\ref{eq:2}) a reasonable inference is
\begin{eqnarray}
{\tensor{P}_{JJ}}(Q)&=& 2\sum_K\frac{\partial\xi_{\textbf{k}+\textbf{q}/2}}
{\partial\textbf{k}}\frac{\partial\xi_{\textbf{k}+\textbf{q}/2}}
{\partial\textbf{k}}\Big[G_KG_{K+Q}\nonumber\\
&-&\Delta_{pg}^2G_{0,-K-Q}^{\gamma} G_{0,-K}^{\gamma}
G_{K+Q}G_K\Big].
\label{eq:3}
\end{eqnarray}


This expression
can be written in a more suggestive notation as
\begin{eqnarray}
2\displaystyle{\sum_K}\frac{\partial\xi_{\textbf{k}+\textbf{q}/2}}{\partial\bf k}\frac{\partial\xi_{\textbf{k}+\textbf{q}/2}}{\partial\bf k}\Bigg(G_{K}G_{K+Q}-F_{pg,K}F_{pg,K+Q}\Bigg)
\label{eq:4}
\end{eqnarray}
with
$F_{pg,K}\equiv\Delta_{pg,\textbf{k}}G^{\gamma}_{0,-K}G_K$.
Interestingly, we have found a similar result for the local density of states
in an STM-based experiment \cite{ourqpi}, but with a different sign in front of the ${pg}$
contribution.

Thus far, we have discussed the current-current correlation function. The
density-density correlation function should necessarily have the same
electromagnetic-vertex function structure which leads to a generalized
particle-hole susceptibility 
\be
& &P_{\rho \rho} (\omega,\vq)=\sumk\int\frac{d\epsilon_1d\epsilon_2}{2\pi^2}
\frac{f(\epsilon_2)-f(\epsilon_1)}{\omega-(\epsilon_1-\epsilon_2)+i\delta}\nonumber\\
&\times&\Big[A_G(\vk+\vq,\epsilon_1)A_G(\vk,\epsilon_2)
+A_{F}(\vk+\vq,\epsilon_1)A_{F}(\vk,\epsilon_2)\Big].\nonumber\\
\nonumber
\ee
Here $A_G$ and $A_{F}$ are the spectral functions for $G$ and
$F_{pg}$.
This expression for $P_{\rho,\rho}$ can only be generalized below $T_c$ by
including the important
contribution from collective modes, often omitted in
the literature \cite{RIXStheory}. Above $T_c$, it is complete.

The sum rules on the longitudinal (L) and transverse (T) components of the
current-current correlation function are a central constraint on our ansatz.
These are given by:
\begin{eqnarray}
\label{eq:6}
\int^{+\infty}_{-\infty}\frac{d\omega}{\pi}\big(-\frac{\textrm{Im}P_{JJ}^L(\omega,\mathbf{q})}{\omega}\big)
&=&\frac{n}{m} \nonumber \\
\textrm{and}~~~ \lim_{q\rightarrow0}\int_{-\infty}^{+\infty}\frac{d\omega}{\pi}\big(-\frac{\textrm{Im}
P_{JJ}^T(\omega,\mathbf{q}) }{\omega}\big)&=&\frac{n}{m}.
\end{eqnarray}

\begin{figure}
\includegraphics[width=3.2in,clip]
{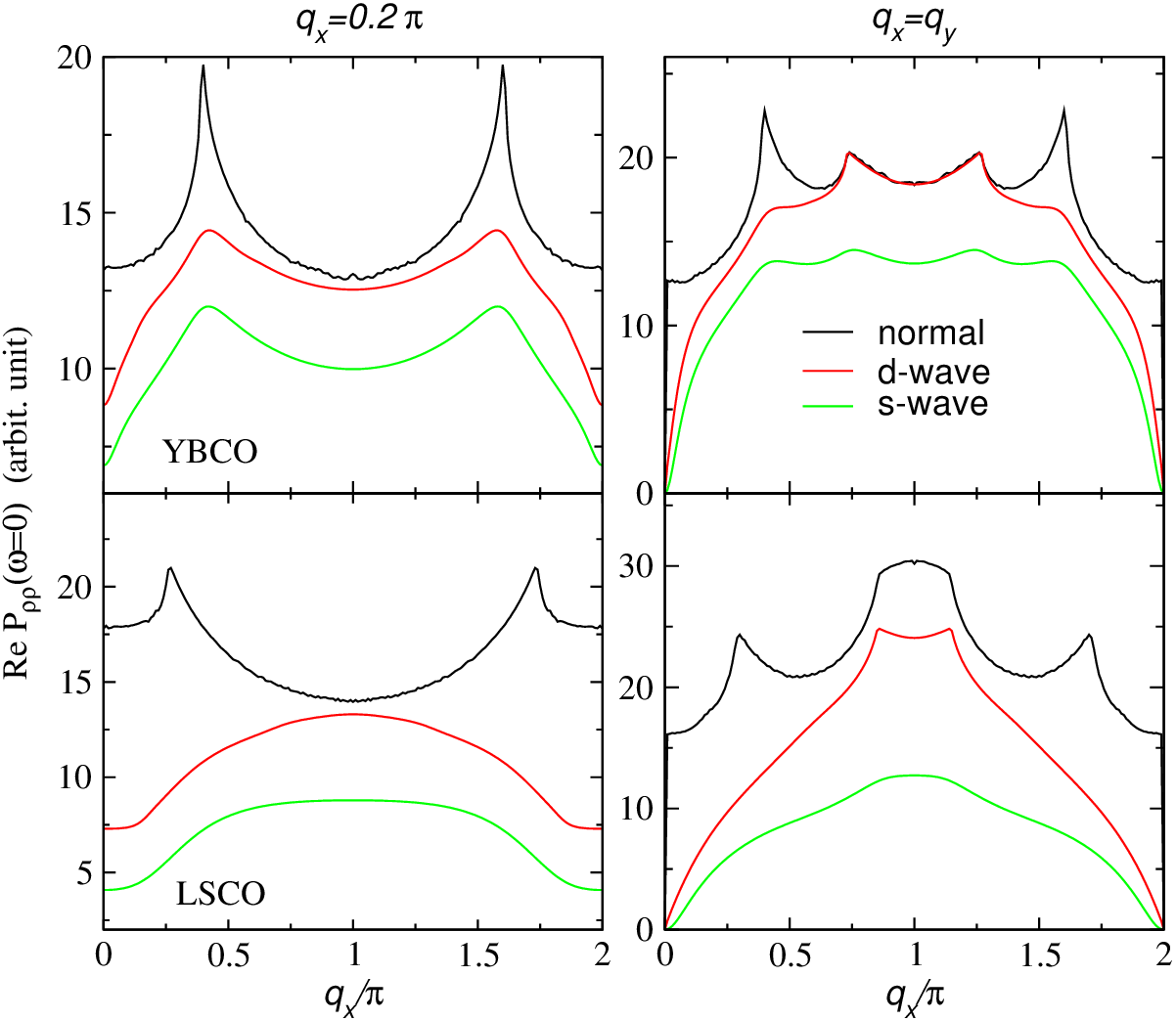}
\caption{Static studies:
Vertical and diagonal cuts plotting Re$P_{\rho,\rho} (\omega=0)$
vs $\mathbf{q}$ for YBCO
and LSCO. The black, red and green lines are for normal gas, $d$-wave and $s$-wave
pseudogaps, respectively. The $d$-wave gap is compatible with nodal peaks but suppresses anti-nodal peaks.
Here $\gamma \approx 0$. }
\label{cut-Re}
\end{figure}

\begin{figure*}
\includegraphics[width=5.8in,clip]
{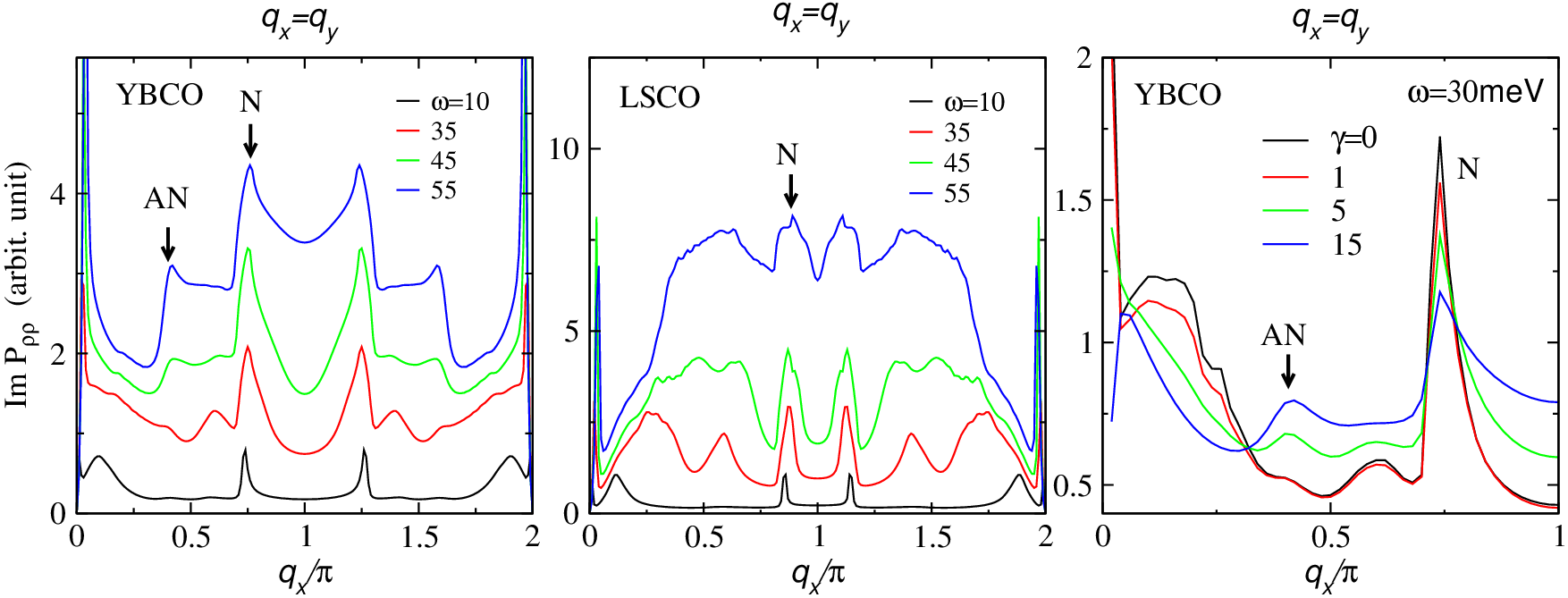}
\caption{Dynamical studies: Left and Central
Figures show diagonal cuts of Im$P_{\rho,\rho}$ vs $\mathbf{q}$ for $d$-wave with YBCO
and LSCO bandstructure, and $\gamma \approx 0$.
The peaks located at $q_x=q_y\approx0.75\pi$ and $q_x=q_y\approx0.4\pi$
become visible between $10-~35~$meV, and correspond
to the nodal (N) and anti-nodal (AN) peaks we observed in Re$P_{\rho,\rho}$
of the
(gapless) normal gas.
There is also a small peak which appears at $q_x=q_y\approx
0.6\pi$ in the $d$-wave curves for $\omega\approx\Delta$.
The plot on the right shows the effect (for YBCO) of varying $\gamma$ at fixed
frequency where one sees
that larger $\gamma$ enhances
the anti-nodal $q_x\approx0.4\pi$ peak
while decreasing the relative height of
the nodal $q_x\approx0.75\pi$ peak.}
\label{cut-Im}
\end{figure*}
Importantly, these sum rules
can be analytically proved using Eq.~(\ref{eq:1}) and
Eq.~(\ref{eq:3}).  The second of these is the weaker condition, as it
provides no real check on the finite $Q$ behavior of the
correlation functions. Indeed, above $T_c$ it can be thought of as an
equivalent condition to the requirement that there is no
Meissner effect. The longitudinal sum rule
represents a more stringent test and is equivalent to
proving a current conservation condition.

To see this \cite{Note1}, note that
the electromagnetic vertex function can be extracted from
the ansatz in Eq.~(\ref{eq:3}).
This vertex satisfies
$\Gamma^{\mu}(K+Q,K)- \gamma^{\mu}(K+Q,K)$
\begin{equation}
=\Delta_{pg}^2\gamma^{\mu}(-K-Q,-K)G^{\gamma}_0(-K-Q)G^{\gamma}_0(-K)
\end{equation}
which is consistent \cite{Note1}
with the Ward Identity $q_{\mu}\Gamma^{\mu}(K+Q,K)=G^{-1}(K+Q)-G^{-1}(K)$.
With this full vertex and
Ward identity, one can verify that the correlation functions
satisfy the current conservation
condition $q^{\mu}Q^{\mu\nu}=0$, so that, for example
\be
\Omega {\bf Q}_{\rho J}-\vq\cdot\tensor{Q}_{JJ}=0.
\label{WI}
\ee
Here
$\tensor{Q}_{JJ} \equiv\tensor{P}+\frac{\tensor{n}}{m}$,
and ${\bf Q}_{\rho J} \equiv 2\sum_K\frac{\p\xi_{\vk+\vq/2}}{\p\vk}G(K+Q)G(K)$.
Using Eq.~(\ref{WI}),
with $\Omega=0$, we find
$\frac{\vq\cdot\tensor{Q}_{JJ}(0,\vq)\cdot\vq}{q^2}=P^L_{JJ}(0,\vq)+\frac{n}{m}=0$.
It then follows that $P^L_{JJ}(0,\mathbf{q})=\int^{+\infty}_{-\infty}\frac{d\omega}{\pi}
\frac{\textrm{Im}P^L_{JJ}(\omega,\mathbf{q})}{\omega}=-\frac{n}{m}$.
which proves consistency between
Eq.~(\ref{eq:3})
and the longitudinal sum rule.

\vskip5mm
\textit{Quantum Oscillations}
To address the theory of quantum oscillations,
we make use of the fact that specific heat data \cite{Boebinger11}
suggest that even in the high magnetic fields the pseudogap
persists. Moreover, we have earlier
shown \cite{Peter2} from Gor'kov theory that at high
fields, when there is only intra-Landau level pairing
\cite{DT},
a BCS-like dispersion persists.
Notably,
the gap or pseudogap parameter is inhomogeneous, but for
some purposes
\cite{MStephen}, this inhomogeneity can be averaged over
in a vortex liquid or a pseudogap phase.
A major effect
of non-zero field is to replace the dispersion $\xi_{\mathbf{p}}$
by the appropriate Landau level quantization.
With this replacement, one can compute an extension of the usual
Lifshitz-Kosevich (LK) formula (based on Eq.~(\ref{eq:1}))
to arrive at an analytic formula
for the density of states as a function of magnetic
field.
The density of states at the Fermi energy is then given by
$N(0)=\frac{H}{(2\pi)^2}\sum_{n,k_z}\frac{\gamma}{\pi(E_{n,k_z}^2+\gamma^2)}$ with $E_{n,k_z}=\sqrt{\xi_{n,k_z}^2+\Delta^2}$
and $\xi_{n,k_z}=(n+\frac12)\omega_c+\frac{k_z^2}{2m}-\mu$,
with $\omega_c=eH/m$.
Using
the Poisson sumation formula, one finds a simple (for the $s$-wave case)
analytic expression for
the oscillatory contribution which depends on non-zero $\gamma$.
A similar analysis follows for the $d$-wave case, although the
result is less compact.

\textit{Numerical Results}
Henceforward, in order not to have too many distinct parameters
we take $\gamma' = \gamma$, although our qualitative findings are
robust for general $\gamma'$.
%
To begin with, in
addressing numerics one can gain analytical intuition
by first considering the limit in which $\gamma =0$
\begin{eqnarray}
&&P_{\rho,\rho}(\textbf{q},\omega)=\sum_{\vk}
\Bigg[\Big(1-\frac{\xi^+\xi^-+\Delta_{pg}^2}{E^+E^-}\Big)\nonumber\\
&&\qquad\times\frac{(E^++E^-)(1-f_+-f_-)}{\omega^2-(E^++E^-)^2}\nonumber\\
&&-\Big(1+\frac{\xi^+\xi^-+\Delta_{pg}^2}{E^+E^-}\Big)
\frac{(E^+-E^-)(f_+-f_-)}{\omega^2-(E^+-E^-)^2}\Bigg]. 
\end{eqnarray}
Here $E_{\pm}=E_{\mathbf{k}\pm\mathbf{q}/2}$,
$\xi_{\pm}=\xi_{\mathbf{k}\pm\mathbf{q}/2}$ and
$f_{\pm}=f(E_{\pm})$.

Importantly, this density response consists of a scattering term
in the third line and (in the second line)
a pair breaking or pair forming term
involving $1 - 2 f$. At the lowest temperatures, the pair breaking
term dominates the spectral weight. Thus the particle-hole response
of a low $T$ system with a pseudogap is only possible when
pairs are broken.

In Figures \ref{cut-Re} and \ref{cut-Im}, we plot the real and imaginary parts of the 
susceptibility $P_{\rho,\rho}(\omega,\vq)$ with the band structure
$\xi_k=t_0+t_1(\cos k_x+\cos k_y)/2+t_2\cos k_x\cos k_y
+t_3(\cos 2k_x+\cos 2k_y)/2$.
Since cuprate bandstructures are somewhat variable \cite{SiLevin}
we consider two different parameter sets.  For YBCO we take:
$t_0=160\mbox{meV},~~~ t_1=-600\mbox{meV},
              ~~~t_2=200\mbox{meV},~~~ t_3=-80\mbox{meV}$.
For LSCO we take
$t_0=130\mbox{meV},~~~ t_1=-600\mbox{meV},
               ~~~t_2=160\mbox{meV},~~~t_3=0\mbox{meV}$.
This yields a square shaped Fermi surface for YBCO and
a rounded shape Fermi surface for LSCO.
We assume the $d$-wave pairing gap is
$\Delta_k=\Delta_0(\cos k_x-\cos k_y)/2$, with
$\Delta_0=35$meV, for definiteness.

In Figure \ref{FS}, we plot (from left to right) the Fermi surfaces of
a normal state YBCO and normal state LSCO system.
It should be clear that the preferred nesting is more antinodal in
YBCO, while more nodal in LSCO.
The lower figure shows quantum oscillations in YBCO
via a plot of the density of states at the Fermi energy
as a function of frequency in the pseudogap phase.
Important
here is the fact that non-zero $\gamma$ (representing the
dynamic equilibrium between pairs and fermions)
enables these oscillations in the presence
of a pseudogap.

Figure \ref{cut-Re} presents a study of the real part of the density-density correlation
function in the static limit. The maxima in this function are generally
associated with
a static, i.e., true, instability of the charge disordered phase.
These plots represent varying $\mathbf{q}$ along
the vertical (left column) as well as diagonal (right column) directions in
Re$P_{\rho,\rho} (\mathbf{q}, \omega=0)$.
The upper panel corresponds to YBCO and the lower to LSCO. Going from
top to bottom (black, red and green) indicates the behavior for the
gapless normal phase, and for the $d$- and $s$-wave paired states.
The peaks for the gapless normal phase in the
left panels represent the antinodal nestings
and they are more apparent for YBCO. The peaks in the gapless normal phase
on the right include nodal nesting and this tends to dominate in
LSCO.  An important observation is that static $d$-wave (or $s$-wave) pairing
is highly destructive to the antinodal peak, whereas the nodal peak
(particularly in YBCO) is very little affected by the $d$-wave pairing gap.
In some respects this seems rather straightforward, and such competition between
pairing and charge ordering in the same regime of $\mathbf{k}$ space
has been discussed much earlier \cite{CunninghamLevin}.
Nevertheless, this underlines the strong competition between a $d$-wave
pseudogap and static antinodal charge ordering.

In Figure \ref{cut-Im}, plots are presented for the behavior of the dynamic charge
susceptibility in the presence of a $d$-wave
pseudogap, for diagonal cuts and a range of frequencies.
The figure on the left represents
YBCO (with $\gamma \approx 0 $), in the center, LSCO, while the
figure on the right shows the effect of variable $\gamma$ in the YBCO case.
In YBCO, the antinodal (AN) peak appears somewhere between $\omega = 10~$meV,
and $\omega = 35~\mbox{meV}= \Delta_0$, becoming
more apparent as frequency increases. (The broad feature below the nodal
maximum (N) in LSCO is not a true anti-nodal peak.) 
At intermediate
$\omega$, over a narrow range, a new peak appears, midway between, 
and reflecting a mixture of the nodal and anti-nodal peaks.

Important to the physical picture are the plots in the right-most
panel showing $\textrm{Im}~P_{\rho,\rho}$ vs.~$\mathbf{q}$
at $\omega = 30~$meV with varying $\gamma$. The figure
demonstrates that as $\gamma$ increases the anti-nodal peak
becomes relatively more important.  Physically, bigger $\gamma$
can be interpreted as reflecting shorter lived pairs. That is, the size of
$\gamma$
reflects the ease with which the finite-lived pairs break up into
their separate fermionic components.
We see in this figure that increasing $\omega$ also assists in
breaking pairs, thus enabling coexistence of dynamic anti-nodal
charge ordering with a $d$-wave pseudogap.

\textit{Conclusions}
The starting point for this paper is Eq.~(\ref{eq:1}) which was derived
from a microscopic t-matrix scheme \cite{Malypapers} independent of
later ARPES phenomenological arguments \cite{Normanphenom}.
Using Gor'kov theory,
we find \cite{Peter2}  Eq.~(\ref{eq:1}) is valid in the presence of a pseudogap in
a very high magnetic field (albeit with $\gamma$ and $\Delta_{pg}$ dependent on $H$).
Our microscopic model \cite{Malypapers} was based on a particular
form for the t-matrix (naturally associated with Gor'kov theory, which
involves one bare and one dressed Green's function). Importantly because
of a gap in the fermionic spectrum, this
form leads to long lived pairs
and a two-peaked spectral function,
thereby distinguishing  it from other (3-peaked) models in the literature
\cite{Micklitz,SenthilLee,Randeria13}.
A crucial finding here is that $H \neq 0$
quantum oscillations persist in a \textit{non}-Fermi liquid phase.

We conclude quite generally that
the pseudogap-phase-derived pairbreaking
through the
parameter $\gamma$, enables the underlying LDA-based fermiology to
be revealed.
Importantly, at finite $\omega$, coexistence of anti-nodal fluctuating order
and a $d$-wave pseudogap becomes possible.
That is, the nesting vectors seen in Figure 1 are evident in
the pseudogap state with non-zero $\omega$ and $\gamma$.
This underlying fermiology was seen in Fermi arcs \cite{ourarpes}
and we have found it
here for charge fluctuations and quantum oscillations.
For the former we have shown that static nodal
order coexists more readily with $d$-wave pairing, while
anti-nodal ordering is more problematic.
We speculate that finite, large $H$ plays
a similar role as $\omega$ and $\gamma$ in enabling, through the
breaking of metastable pairs, the coexistence of
(in this case) a static antinodal charge ordering and a $d$-wave pseudogap,
as observed \cite{Hardy13}.

\vskip3mm
Work supported by NSF-MRSEC Grant
0820054. P.S.~acknowledges support from the Hertz Foundation.


\end{document}